\documentclass[twocolumn,amsmath,amssymb,prl,aps]{revtex4}
\input psfig
\begin{document}

\title{What do we (not) know theoretically about solar neutrino fluxes?}
\author{John N. Bahcall\thanks{E-mail: jnb@ias.edu}}
\affiliation{School of Natural Sciences, Institute for Advanced Study, Princeton, New Jersey 08540}
\author{M. H. Pinsonneault
\thanks{E-mail: pinsono@payne.mps.ohio-state.edu}}
\affiliation{Department of Astronomy, Ohio State University, Columbus, OH 43210}

\begin{abstract}Solar model
predictions of  $^8$B  and   p-p neutrinos agree with the
experimentally-determined fluxes (including oscillations):
$\phi({\rm pp})_{\rm measured} =(1.02 \pm 0.02 \pm 0.01)\phi({\rm
pp})_{\rm theory} $, and $\phi({^8{\rm B}})_{\rm measured} =(0.88
\pm 0.04 \pm 0.23) \phi({^8{\rm B}})_{\rm theory}$, $1\sigma$
experimental and theoretical uncertainties, respectively. We use
improved input data for nuclear fusion reactions, the equation of
state, and the chemical composition of the Sun. The solar
composition is the dominant uncertainty in calculating the $^8$B
and CNO neutrino fluxes; the cross section for the
$^3$He($^4$He,$\gamma$)$^7$Be reaction is the most important
uncertainty for the calculated $^7$Be neutrino flux.
\end{abstract}

\pacs{Valid PACS appear here} \maketitle
\maketitle

This paper is part of a series that spans more than 40 years~\cite{series}. The goals of this series are
to provide increasingly more precise theoretical calculations of  the solar neutrino fluxes and detection
rates and to make increasingly more comprehensive evaluations of the uncertainties in the predictions. We
describe here two steps forward (improved accuracy of the equation of state of the solar interior and
some of the  nuclear fusion  data) and one step backward (increased systematic uncertainties in the
determination of the surface composition of the Sun).

Using recent improvements in input data, we calculate the best-estimates, and especially the
uncertainties, in the solar model predictions of solar neutrino fluxes. We compare the calculated
neutrino fluxes with their measured values.We stress the need for more accurate measurements of the
surface composition of the Sun and of specific nuclear reaction rates.

\begin{table}[!t]
\centering \caption[]{Predicted solar neutrino fluxes from solar
models. The table presents the predicted fluxes, in units of
$10^{10}(pp)$, $10^{9}({\rm \, ^7Be})$, $10^{8}(pep, {\rm ^{13}N,
^{15}O})$, $10^{6} ({\rm \, ^8B, ^{17}F})$, and $10^{3}(hep)$
${\rm cm^{-2}s^{-1}}$. Columns 2-4 show BP04, BP04+, and our
previous best model BP00~\cite{bp00}. Columns 5-7 present the
calculated fluxes for solar models that differ from  BP00  by an
improvement in one set of input data: nuclear fusion cross
sections (column 5), equation of state for the solar interior
(column 6), and surface chemical composition for the Sun (column
7). Column~8  uses the same input data as for BP04 except for a
recent report of the $^{14}$N + p fusion cross section. References
to the improved input data are given in the text. We use OPAL
radiative opacities  calculated for each chemical composition. The
last two rows ignore neutrino oscillations and present for the
chlorine and gallium solar neutrino experiments the capture rates
in SNU (1 SNU equals $10^{-36}{\rm ~events
~per~target~atom~per~sec}$). Due to oscillations, the measured
rates are smaller: $2.6 \pm 0.2$ and $69 \pm 4$, respectively. We
use the neutrino absorption cross sections and their uncertainties
that are given in Ref.~\cite{nucrosssections}.
 \protect\label{tab:neutrinofluxes}}
\begin{tabular}{lccccccc}
\noalign{\bigskip} \hline\hline \noalign{\smallskip}
Source&\multicolumn{1}{c}{BP04}&{BP04+}&BP00&Nucl&EOS&Comp&$^{14}$N\\
\noalign{\smallskip} \hline \noalign{\smallskip}
$pp$&5.94$(1 \pm 0.01)$&5.99 &5.95&5.94&5.95&6.00&5.98\\
$pep$&1.40$(1 \pm 0.02) $&1.42&1.40&1.40&1.40&1.42&1.42\\
$hep$&$7.88 (1 \pm 0.16)$&8.04&9.24&7.88&9.23&9.44&7.93\\
${\rm ^7Be}$&$4.86 (1 \pm 0.12)$&4.65&4.77&4.84&4.79&4.56&4.86\\
${\rm ^8B}$&5.79$(1 \pm 0.23)$&5.26&5.05&5.77&5.08&4.62&5.74\\
${\rm ^{13}N}$&$5.71(1~~^{+0.37}_{-0.35}) $&4.06&5.48&5.69&5.51&3.88&3.23\\
${\rm ^{15}O}$&$5.03(1~~^{+0.43}_{-0.39}) $&3.54 &4.80&5.01&4.82&3.36&2.54\\
${\rm ^{17}F}$&$5.91(1~~^{+0.44}_{-0.44}) $&3.97&5.63&5.88&5.66&3.77&5.85\\
\noalign{\smallskip} \hline \noalign{\smallskip}
Cl&$8.5^{+1.8}_{-1.8}$&7.7&7.6&8.5&7.6&6.9&8.2\\
Ga&131$^{+12}_{-10}$&126&128&130&129&123&127\\
\noalign{\smallskip} \hline\hline \noalign{\smallskip}
 \noalign{\smallskip}
\end{tabular}
\end{table}

Table~\ref{tab:neutrinofluxes} presents, in the second   (third)
column, labeled BP04 (BP04+), our best solar model calculations
for the neutrino fluxes. The uncertainties are given in column~2.
BP04+ was calculated with new input data for the equation of
state~\cite{eos}, nuclear physics~\cite{b8,pphep}, and solar
composition~\cite{newcomp}. BP04, our currently preferred model,
is the same as BP04+ except that BP04 does not include the most
recent analyses of the solar surface composition~\cite{newcomp},
which conflict with helioseismological measurements. The error
estimates, which are the same for BP04, BP04+, and $^{14}$N (see
Table~\ref{tab:neutrinofluxes}), include the recent composition
analyses.

For the BP04 solar model, the base (mass) of the convective zone
is $0.715 R_\odot$ (0.024 $M_\odot$), the surface heavy element to
hydrogen ratio by mass, $Z/X = 0.0229$,  the surface helium
abundance is 0.243, and 1.6\% of the luminosity is from CNO
reactions. The central temperature, helium abundance, and Z/X,
are, respectively, $15.72 \times 10^6\ $K, 0.640, and 0.0583. All
of these values are in the acceptable range as determined by
helioseismology. However, for BP04+, the base of the convective
zone (CZ) is $R_{\rm CZ}/R_\odot = 0.726$, which conflicts with
the measured value of $0.713 \pm 0.001$ (or, $\pm 0.003$, see
Ref.~\cite{cz}).  By examining a series of models, we have
determined that the reason for the too-shallow CZ in the BP04+
model is the lower heavy element abundance, $Z/X = 0.0176$.
Therefore, we prefer BP04.

The  measurements from different solar neutrino
experiments~\cite{nudata} and the KamLAND reactor
data~\cite{kamland} can be combined in a global analysis to obtain
the best empirical values for the p-p, $^8$B, and $^7$Be solar
neutrino fluxes. We use  the fluxes from  the global analysis of
Ref.~\cite{lowerenergy},  which allows all the solar neutrino
fluxes to be free parameters subject only to the luminosity
constraint (i.e., energy conservation). Comparing the measured
values with the theoretical predictions, we find for BP04:

\begin{eqnarray}
\phi({\rm pp})_{\rm measured} &=&(1.02 \pm 0.02 \pm 0.01)\phi({\rm pp})_{\rm theory} \label{eqn:pp} \\
\phi({^8{\rm B}})_{\rm measured} &=&(0.88 \pm 0.04 \pm 0.23) \phi({^8{\rm B}})_{\rm theory} \label{eqn:8b} \\
\phi({^7{\rm Be}})_{\rm measured} & = & (0.91^{+0.24}_{-0.62} \pm 0.11)\phi({^7{\rm Be}})_{\rm
theory}
\label{eqn:7be}
\end{eqnarray}
In Eq.~(\ref{eqn:pp})-Eq.~(\ref{eqn:7be}), the $1\sigma$ experimental uncertainties are given before the
$1\sigma$ theoretical uncertainties.

The  measured and  theoretical values for the fluxes agree within their combined $1\sigma$ uncertainties.
The measurement error of the $^8$B neutrino flux is smaller than the uncertainty in the theoretical
calculation, but the opposite is true for the p-p and $^7$Be neutrino fluxes.

Column four of Table~\ref{tab:neutrinofluxes} presents the fluxes
calculated using our previous best solar model, BP00~\cite{bp00}.
The BP04 best-estimate neutrino fluxes and their uncertainties
have not changed markedly from their BP00 values despite
refinements in input parameters. The only exception is the CNO
flux uncertainties which have almost doubled   due to the larger
systematic uncertainty in the surface chemical composition
estimated in this paper.

We describe   improvements in the input data relative to BP00. Quantities that are not discussed here are
the same as for BP00. Each class of improvement is represented by a separate column, columns 5-7,  in
Table~\ref{tab:neutrinofluxes}.

Column~5 contains the fluxes computed for a solar model that is
identical to BP00 except that we have used improved values for
direct measurements of the $^7$Be(p,$\gamma$)$^8$B cross section,
$S_{\rm 20~keV}(^7{\rm Be + p}) = 20.6 \pm 0.8~{\rm
eV~b}$~\cite{b8}, and the calculated p-p, $S_0(pp) = 3.94(1 \pm
0.004)\times 10^{-25}\, {\rm MeV ~ b}$, and hep, $S_0({\rm hep}) =
( 8.6 \pm 1.3) \times 10^{-20}\, {\rm keV ~ b}$, cross
sections~\cite{pphep}. The reactions that produce the $^8$B and
hep neutrinos are rare; changes in their production cross sections
only affect, respectively, the $^8$B and hep fluxes. The 15\%
increase in the calculated $^8$B neutrino flux, which is primarily
due to a more accurate cross section for $^7$Be(p,$\gamma$)$^8$B,
is the only significant change in the best-estimate fluxes.

The fluxes in Column~6 were calculated using a refined equation of
state, which includes relativistic corrections and a more accurate
treatment of molecules~\cite{eos}. The equation of state
improvements between 1996 and 2001, while significant in some
regions of parameter space, change all the solar neutrino fluxes
by less than 1\%. Solar neutrino calculations are insensitive to
the present level of uncertainties in the equation of state.

The most important changes in the astronomical data since BP00
result from new analyses of the surface chemical composition of
the Sun. The input chemical composition affects the radiative
opacity and hence the physical characteristics of the solar model,
and to a lesser extent the nuclear reaction rates. New values for
C,N,O,Ne, and Ar have been derived~\cite{newcomp} using
three-dimensional rather than one-dimensional atmospheric models,
including hydrodynamical effects, and paying particular attention
to uncertainties in atomic data and observational spectra. The new
abundance estimates, together with the previous best-estimates for
other solar surface abundances~\cite{oldcomp}, imply $Z/X =
0.0176$, much less than the previous value of $Z/X =
0.0229$~\cite{oldcomp}. Column~7 gives the fluxes calculated for
this new composition mixture. The largest change in the neutrino
fluxes for the  p-p chain is the 9\% decrease   in the predicted
$^8$B neutrino flux. The N and O fluxes are decreased by much
more, $\sim 35 \%$, because they reflect directly the inferred C
and O abundances.

The CNO nuclear reaction rates are less well determined than the
rates for the more important (in the Sun) p-p
reactions~\cite{adelberger}. The rate for
$^{14}$N(p,$\gamma$)$^{15}$O is poorly known, but important for
calculating CNO neutrino fluxes.  Extrapolating to the low
energies relevant for solar fusion introduces a large uncertainty.
Column 8 gives the neutrino fluxes calculated with input data
identical to BP04  except for the  cross section factor $S_0({\rm
^{14}N + p}) = 1.77 \pm 0.2\, {\rm keV~b}$ that is about half the
current best-estimate;  this value assumes a particular R-matrix
fit to the experimental data~\cite{new14n}. The p-p cycle fluxes
are changed by only $\sim 1$\%, but the $^{13}$N and $^{15}$O
neutrino fluxes are reduced by $ 40\%-50$\% relative to the BP04
predictions. CNO nuclear reactions contribute 1.6\% of the solar
luminosity in the BP04 model and only 0.8\% in the model with a
reduced $S_0({\rm ^{14}N + p})$.

Table~\ref{tab:uncertainties} shows the individual contributions to  the flux uncertainties.  These
 uncertainties  fix the accuracy to which a given input parameter should be determined.
Improvements in input data enable  more precise tests of stellar evolution via solar neutrino
experiments.

Columns~2-5 present the fractional uncertainties from the nuclear reactions whose measurement errors are
most important for calculating neutrino fluxes. Unless stated otherwise, we have used throughout this
paper the uncertainties estimated in Ref.~\cite{adelberger} for nuclear cross sections.

The measured rate of the $^3$He-$^3$He reaction, which after the inception of this series~\cite{series}
changed by a factor of 4, and the measured rate of the $^7$Be + p reaction, which for most of this series
has been the dominant uncertainty in predicting the $^8$B neutrino flux, are by  now very well
determined. If the published systematic uncertainties for the $^3$He-$^3$He and $^7$Be + p reactions are
correct,then their uncertainties no longer contribute in a crucial way to the calculated theoretical
uncertainties (see column~2 and column~4 of Table~\ref{tab:uncertainties}). These important improvements
in our knowledge of low energy nuclear cross section factors have been achieved by increasingly precise,
but always enormously difficult and beautiful, experiments carried out over four decades by many
researchers in laboratories throughout the world.

\begin{table}[!t]
\centering \caption[]{Principal sources of uncertainties in calculating solar neutrino fluxes.
 Columns 2-5 present the fractional uncertainties in the neutrino fluxes from laboratory
 measurements of, respectively, the $^3$He-$^3$He, $^3$He-$^4$He, p-$^7$Be, and p-$^{14}$N
 nuclear fusion reactions. The last four columns, 6-9, give, respectively, the fractional uncertainties
 due to  the calculated radiative opacity, the calculated rate of element diffusion,
 the measured solar luminosity, and the measured heavy element to hydrogen ratio.\protect\label{tab:uncertainties}}
\begin{tabular}{lccccccccc}
\noalign{\bigskip} \hline\hline \noalign{\smallskip}
Source&\multicolumn{1}{c}{3-3}&3-4&1-7&1-14&Opac&Diff&$L\odot$&Z/X\\
\noalign{\smallskip} \hline \noalign{\smallskip}
$pp$&0.002&0.005&0.000&0.002&0.003&0.003&0.003&0.010\\
$pep$&0.003&0.007&0.000&0.002&0.005&0.004&0.003&0.020\\
$hep$&0.024&0.007&0.000&0.001&0.011&0.007&0.000&0.026\\
${\rm ^7Be}$&0.023&0.080&0.000&0.000&0.028&0.018&0.014&0.080\\
${\rm ^8B}$&0.021&0.075&0.038&0.001&0.052&0.040&0.028&0.200\\
${\rm ^{13}N}$&0.001&0.004&0.000&0.118&0.033&0.051&0.021&0.332\\
${\rm ^{15}O}$&0.001&0.004&0.000&0.143&0.041&0.055&0.024&0.375\\
${\rm ^{17}F}$&0.001&0.004&0.000&0.001&0.043&0.057&0.026&0.391\\
\noalign{\smallskip} \hline\hline \noalign{\smallskip}
 \noalign{\smallskip}
\end{tabular}
\end{table}

The most important nuclear physics uncertainty in calculating
solar neutrino fluxes is now the rate of the $^3$He-$^4$He
reaction (column~3 of Table~\ref{tab:uncertainties}).  The
systematic uncertainty in the the rate of $^3$He($^4$He,
$\gamma$)$^7$Be reaction(see Ref.~\cite{adelberger})  causes an
8\% uncertainty in the prediction of both the $^7$Be and the $^8$B
solar neutrino fluxes. It is astonishing that there has not been
any progress in the past 15 years in measuring this rate more
accurately.

For $^{14}$N(p,$\gamma$)$^{15}$O, we have continued to use in Table~\ref{tab:uncertainties} the
uncertainty given in Ref.~\cite{adelberger}, although the recent reevaluation in Ref.~\cite{new14n}
suggests that the uncertainty could be somewhat larger (see column~7 of Table~\ref{tab:neutrinofluxes}).

The uncertainties due to the calculated radiative opacity and element diffusion, as well as the measured
solar luminosity (columns 6-8 of Table~\ref{tab:uncertainties}), are all moderate, non-negligible but not
dominant. For the $^8$B and CNO neutrino fluxes, the uncertainties that are due to the radiative opacity,
diffusion coefficient, and solar luminosity are all in the range 2\% to 6\%.

The surface composition of the Sun is the most problematic and
important source of uncertainties. Systematic errors dominate: the
effects of line blending, departures from local thermodynamic
equilibrium, and details of the model of the solar atmosphere. We
assume that the uncertainty in all important abundances is
approximately the same. We have defined previously the $3\sigma$
range of $Z/X$ as the spread over all modern determinations (see
Refs.~\cite{series,bp00,book}), which now implies $\Delta
(Z/X)/(Z/X) = 0.15 ~(1\sigma)$, 2.5 times larger than the
uncertainty adopted in BP00. The most recent uncertainty quoted
for oxygen, the most abundant heavy element in the Sun, is
similar: 12\% \cite{newcomp}.

Heavier elements like Fe affect the radiative opacity and hence
the neutrino fluxes more strongly than the relatively light
elements~\cite{bp00}.  This is the reason why the difference
between the fluxes calculated with BP04 and BP04+ (or between BP00
and Comp, see Table~\ref{tab:neutrinofluxes}) is less than would
be expected for the 23\% decrease in $ Z/X$. The abundances that
have changed significantly since BP00 (C, N, O, Ne, Ar) are all
for lighter species for which meteoritic data are not available.

 The dominant uncertainty listed in Table~\ref{tab:uncertainties} for
the $^8$B and CNO neutrinos is  the chemical composition,
represented by $Z/X$ (see column~9). The uncertainty ranges from
20\% for the $^8$B neutrino flux to $\sim 35$\% for the CNO
neutrino fluxes. Since the publication of BP00, the best published
estimate for Z/X decreased by $4.3\sigma$(BP00 uncertainty) and
the estimated uncertainty due to $Z/X$ increased for $^8$B
($^{15}$O) neutrinos by a factor of 2.5 (2.8). Over the past three
decades, the changes have almost always been toward a smaller
$Z/X$. The monotonicity  is surprising since different sources of
improvements have caused successive changes. Nevertheless, since
the changes are monotonic, the  uncertainty estimated from the
historical record is large.

We list below our principal conclusions and their implications.
First, the experimentally-determined values for the p-p, $^7$Be,
and $^8$B solar neutrino fluxes are in agreement with the values
predicted by the standard solar model. In each case, the agreement
is accidentally better than the combined $1\sigma$ uncertainties.
More precise measurements of the p-p and $^7$Be fluxes will test
critically the theory of energy generation in the solar interior.
Second, recent precise measurements or  improved calculations of
nuclear reaction parameters, the equation of state, and the
surface chemical composition of the Sun have refined the input
data to the solar model calculations but have not changed the
calculated neutrino fluxes outside the previously-quoted
theoretical $1\sigma$ uncertainties (see columns~3-6 of
Table~\ref{tab:neutrinofluxes} and Ref.~\cite{bp00}). Third, the
rate of the reaction $^3$He($^4$He,$\gamma$)$^7$Be is the largest
nuclear physics contributor to the uncertainties in the solar
model predictions of the  neutrino fluxes in the p-p chain.  In
the past 15 years, no one has remeasured this rate; it should be
the highest priority for nuclear astrophysicists. For the
important $^7$Be neutrino flux that can be measured in the
BOREXINO~\cite{borexino} and KamLAND~\cite{kamland} detectors,
there is currently an 8\% uncertainty due to the cross section for
$^3$He($^4$He,$\gamma$)$^7$Be. Fourth, the cross section for the
reaction $^{14}$N(p,$\gamma$)$^{15}$O is the largest nuclear
physics contributor to the uncertainties in the calculated CNO
neutrino fluxes. It is important to measure this cross section
more accurately in order to understand well energy production in
stars heavier than the Sun. Neutrino oscillation studies can use
the fluxes from both the low $^{14}$N model and BP04 (see
Table~\ref{tab:neutrinofluxes}); the results for oscillation
parameters should be essentially identical. Fifth, the largest
uncertainty in calculating all the solar neutrino fluxes is now
the uncertainty in the measured surface composition of the Sun
(see Table~\ref{tab:uncertainties}). Unfortunately, the principal
uncertainties in inferring the composition are systematic. The
uncertainties could be estimated more reliably, and perhaps
reduced, by comparing the results from a variety of different
measurements of the same elemental abundances and by different
groups of astrophysicists making comparably sophisticated two and
three dimensional hydrodynamical models of the solar outer region.
Sixth, the recent reanalyses~\cite{newcomp} of the solar chemical
composition imply a lower surface heavy element abundance and
consequently a base of the convective zone that conflicts with
helioseismological measurements~\cite{cz}. For this reason, we
have not used the most recent low heavy element abundances in our
preferred model, BP04. The low heavy element abundances lead in
BP04+ to a slightly better agreement with the $^8$B neutrino
measurement, but the improvement between best-estimates is smaller
than the $1\sigma$ uncertainty.

We make three recommendations for future work, based on the known dependences of neutrino fluxes on input
parameters~\cite{book}.  First, the low energy cross section of the $^3$He($^4$He,$\gamma$)$^7$Be
reaction should be measured to better than $\pm 5$\% ($1\sigma$)(a factor of two improvement) in order
that the uncertainty in this reaction not limit the interpretation of future $^7$Be solar neutrino
experiments. Second, the uncertainty in the surface heavy element abundances (particularly elements like
iron that contribute most significantly to the radiative opacity~\cite{bp00})  should be reduced to
 less than $\pm 0.02$ dex (a factor of 3 improvement) in order that the calculational uncertainty from
the composition not exceed the current error ($\pm
4$\%~\cite{lowerenergy,nudata,kamland}) in the empirical
determination of the $^8$B neutrino flux. Third, the uncertainty
in the low energy extrapolation of the rate of the
$^{14}$N(p,$\gamma$)$^{15}$O reaction must be $\leq 25$\% in order
that the p-p flux can be used for a precision measurement of the
neutrino mixing angle $\theta_{12}$~\cite{lowerenergy} and an
accurate test of stellar evolution theory.

 \acknowledgments JNB acknowledges  NSF grant No. PHY0070928.

\end{document}